\documentclass[a4paper,twocolumn,aps,prl,nolongbibliography,superscriptaddress,showpacs,showkeys,amsmath,amssymb,floatfix,nofootinbib]{revtex4-1}
\usepackage{lmodern}

\usepackage[T1]{fontenc}
\usepackage[utf8]{inputenc}
\usepackage{color}
\usepackage[unicode=true,pdfusetitle,
 bookmarks=true,bookmarksnumbered=true,bookmarksopen=true,bookmarksopenlevel=1,
 breaklinks=false,pdfborder={0 0 0},backref=false,colorlinks=true]
 {hyperref}
\hypersetup{
 citecolor=blue,filecolor=blue,linkcolor=blue,urlcolor=blue}
\makeatletter

\usepackage{amsmath}
\usepackage{slashed}
\usepackage{braket}
\usepackage{braket}
\pdfpageheight\paperheight
\pdfpagewidth\paperwidth

 \@ifundefined{textcolor}{}
 {%
   \definecolor{BLACK}{gray}{0}
   \definecolor{WHITE}{gray}{1}
   \definecolor{RED}{rgb}{1,0,0}
   \definecolor{green}{rgb}{0,0.7,0}
   \definecolor{BLUE}{rgb}{0,0,1}
   \definecolor{CYAN}{cmyk}{1,0,0,0}
   \definecolor{MAGENTA}{cmyk}{0,1,0,0}
   \definecolor{YELLOW}{cmyk}{0,0,1,0}
 }

\renewcommand{\[}{\begin{equation}}
\renewcommand{\]}{\end{equation}} 
\newcommand{\nn}{\nonumber \\}
\usepackage{array}
\makeatother

\begin{document}

\title{Covariantly Quantum Galileon}

\author{Ippocratis D. Saltas} 
\email{isaltas@fc.ul.pt}
\affiliation{Institute of Astrophysics \& Space Sciences, Faculty of Sciences,\\ Campo Grande, PT1749-016 Lisboa, Portugal}

\author{Vincenzo Vitagliano}
\email{vincenzo.vitagliano@ist.utl.pt}
\affiliation{Multidisciplinary Center for Astrophysics \& Department of Physics,\\ Instituto Superior T\'ecnico, University of Lisbon, Av. Rovisco Pais 1, 1049-001 Lisboa, Portugal}


\begin{abstract}
We derive the 1--loop effective action of the cubic Galileon coupled to quantum--gravitational fluctuations in a background and gauge-independent manner, employing the covariant framework of DeWitt and Vilkovisky. Although the bare action respects shift symmetry, the coupling to gravity induces an effective mass to the scalar, of the order of the cosmological constant, as a direct result of the non--flat field--space metric, the latter ensuring the field--reparametrization invariance of the formalism. 
Within a gauge--invariant regularization scheme, we discover novel, gravitationally induced non--Galileon higher-derivative interactions in the effective action. These terms, previously unnoticed within standard, non-covariant frameworks, are not Planck suppressed. Unless tuned to be sub-dominant, their presence could have important implications for the classical and quantum phenomenology of the theory.
\end{abstract}
\maketitle

\section{Introduction} 
The coupling of gravity to matter fields has been an important area of research from particle physics to cosmology, the simplest and most well--studied case being that of a canonical scalar field minimally coupled to Einstein--Hilbert gravity. In recent years, mostly motivated by the dark-energy problem, the emergence of a new family of theories with non--trivial derivative structure and second-order equations of motion, dubbed Galileon theories \cite{Horndeski:1974wa,Nicolis:2008in,Deffayet:2009wt}, has challenged our understanding of their classical and quantum dynamics coupled or not to gravity. 

The physical significance of the Galileon theories is due to their emergence as effective limits in a diverse set of contexts, e.g higher--dimensional and massive gravity theories \cite{Dvali:2000hr,deRham:2010kj,deRham:2012az}. On the same time, they lead to a rich cosmological phenomenology \cite{Chow:2009fm,Silva:2009km,Kobayashi:2010cm,DeFelice:2010nf}, posses intriguing renormalization properties and exhibit the so--called Vainshtein screening mechanism. The Vainshtein mechanism \cite{Vainshtein:1972sx} relies on non--trivial derivative configurations of the scalar to suppress its effects close to matter sources, and avoid observational discrepancies with General Relativity. What is more, it has been shown (see \cite{Luty} for cubic Galileon and \cite{Hinter} for a later generalization) that loops of the scalar field $\phi$ do not renormalize the Galileon interactions, also when the Galileon is coupled to a heavy scalar field that respects the symmetries of the theory \cite{Hinter, deRham:2014zqa}, a result closely related to the special symmetry Galileon theories enjoy, the so--called Galilean symmetry. 

In this work, we will consider an action for gravity minimally coupled to the leading-order Galileon (cubic) term, given by
\begin{align}
S=\int  d^{4}x \sqrt{-g} \left[ -\frac{2}{\kappa^2} R +X\left(1 + \frac{\Box \phi}{M^3} \right)+\frac{4\Lambda}{\kappa^2}\right], \label{action:0}
\end{align}
with $X \equiv \frac{1}{2}g^{\mu \nu} \nabla_\mu\phi \nabla_\nu \phi$, $\Box \equiv  g^{\mu \nu} \nabla_\mu \nabla_\nu$, $\kappa^2 \equiv 32\pi G$, $R$ the Ricci scalar, $\Lambda$ the cosmological constant, and $M$ an in principle arbitrary mass scale.

The quantum dynamics of these theories in the presence of graviton loops has been so far unknown. In this work, we make a step forward by calculating for the first time the effective action at 1--loop order, including the quantum back-reaction of gravity. We will be working at quadratic order in background fields, hence our result will not include the renormalization of the cubic term itself. However, as we show, the presence of the cubic term will lead to novel, non--trivial effects of gravitational origin in the effective action with consequences on both quantum and classical dynamics of the theory. 

The existence of a screening mechanism that relies on non--trivial configurations of the scalar, as well as the consideration of gravitational fluctuations makes the choice of background and gauge crucial. What is more, the calculation of the effective action usually relies on some assumption about the split between background and fluctuating field variables, in principle leading to a further dependence of the results on the choice of background and gauge \footnote{See refs \cite{Labus:2016lkh,Ohta:2016npm} for a recent discussion of associated issues.}. Arguing that such dependence of the effective action is a by-product of the quantum fields' dependence on the parametrization, Vilkovisky \cite{Vilkovisky:1984st,gospel} and DeWitt \cite{deW} developed a geometrical, field-space covariant procedure to modify the usual background-field method which renders the effective action gauge- and background-independent (see also \cite{Parker:2009uva} for a pedagogical introduction). The Vilkovisky-DeWitt formalism has been employed to investigate purely quantum gravitational contributions to the beta functions for gravity with a cosmological constant \cite{Toms:2008dq} (see also \cite{Donkin:2012ud} for a functional RG approach), and in the context of quantization of nonminimally coupled scalar-tensor theories \cite{Mackay:2009cf, Pietrykowski:2012nc} and Einstein-Maxwell systems \cite{Toms:2010vy, Toms:2009vd}. Very recently, the covariant quantization procedure has been adopted to show how the stability of the Higgs vacuum during inflation constrains the curvature-Higgs coupling \cite{Moss:2015gua}. This is the framework we will be using in this work. An extra complication usually appears through the need to regularize traces of infinite field fluctuations of virtual particles running in loops - in this context, we will be employing the technique of dimensional regularization, which manifestly preserves the gauge structure of the theory.

Our main result, as presented in (\ref{Eff-action}), is the covariant 1--loop effective action for the cubic Galileon accounting for quantum-gravitational corrections.  We show that the covariant approach reveals novel derivative interactions of the Galileon field, of gravitational origin, unseen in standard computations of the effective action. 

In what follows, we gently introduce to the unfamiliar reader the framework of the covariant effective action and then present our main results, highlighting in a final section our main  conclusions. Some useful explicit formulae are presented in the Appendix. 

\section{Covariant 1-loop effective action\label{sec:Cov-action}}

Our starting point is the (Euclidean) 1--loop, field-reparametrization invariant effective action $\Gamma$, 
\begin{align}
\Gamma = -\ln \int [d\eta]e^{- S_{\text{quad}}[\bar{\Phi}; \eta] },  \label{action:effective0}
\end{align}
with the Gaussian piece of the bare action $S$ defined through
\begin{align}
S_{\text{quad}} = \frac{1}{2} \lim_{\alpha \to 0} \eta^i \eta^j \left( \nabla_i \nabla_j S + \frac{1}{2\alpha} K^{\alpha}_{i} K_{j \alpha} \right). \label{eq:S_quad}
\end{align}
Here, $i,j$ are generalized field/spacetime indices, and $\eta^{i} = \{h_{\alpha \beta}, \psi\}$ denotes metric/scalar--field fluctuations 
around some background, i.e $\Phi^i = \bar{\Phi}^i + \eta^i$ \footnote{For simplicity, we will drop the overbar for background quantities from now on.}. $K^{\alpha}_{i}$ are the generators of (infinitesimal) symmetries enjoyed by the theory. Their presence in (\ref{eq:S_quad}) defines the gauge-fixing part of the quadratic action as $K_{i \alpha} \eta^{i} = 0$, with its explicit form to be defined later.

The effective action (\ref{action:effective0}) enjoys a generalized covariance: it is a scalar not only under infinitesimal transformations, but also under general field re-definitions. In this sense, it is covariant in field-space, with the corresponding metric denoted as $\mathfrak{g}_{ij}$. 
The associated covariant functional derivative ($\nabla_i$) is built out of $\mathfrak{g}_{ij}$, and lies in the heart of field-reparametrization invariance. It is evaluated in the usual way as 
\begin{align}
\nabla_i \nabla_j S = \partial_i \partial_j S  - \Gamma^{k}_{ij} \partial_\kappa S, \label{eq:cov-action}
\end{align}
with $\Gamma^{k}_{ij}$ the Christoffel connection built out of $\mathfrak{g}_{ij}$ and $\partial_i$ understood as functional derivatives with respect to the field $\Phi^i$. The Landau--DeWitt gauge choice is recovered in the limit $\alpha \to 0$, and it has been shown to be coinciding with the gauge independent result with any other gauge choice \cite{Fradkin:1983nw, Rebhan:1986wp}. Notice that on-shell ($\partial_\kappa S = 0$), the connection terms vanish, a manifestation of the fact that physical (on--shell) quantities do not depend on the parametrization of the fields.

Consistency of the covariant construction requires the metric of the field-space, $\mathfrak{g}_{ij}$, to be ultralocal (i.e. not containing derivatives of the fields), diagonal and to solve the Killing equation, \begin{equation}
\mathfrak{g}_{ij,k}K^k_\alpha+2K^k_{\alpha,(i}\mathfrak{g}_{j)k}=0, 
\end{equation}
in order for the gauge group to generate isometries in the field-space \cite{Vilkovisky:1984st, Demmel:2015zfa}. According to DeWitt \cite{deW}, there exists a unique metric $\mathfrak{g}_{ij}$ without new dimensionful parameters, whose non-zero elements are
\begin{align}
\mathfrak{g}_{g_{\mu \nu}(x) g_{\rho \sigma}(x')} &= \sqrt{g(x)}\cdot \left( g^{\mu (\rho}g^{\sigma) \nu}  - \frac{1}{2}g^{\mu \nu}g^{\rho \sigma} \right)\delta(x,x'),\nn
\mathfrak{g}_{\phi(x) \phi(x')} &= \sqrt{g(x)}\delta(x,x'),
\end{align}
and the non-zero field-space connection terms are ($d$ is the number of spacetime dimensions) 
\begin{align}
& 
 \Gamma^{g_{\mu \nu}(x)}_{\phi(x') \phi(x'')}=\frac{1}{2(d-2)}g_{\mu\nu}\delta(x,x')\delta(x',x''),\nn  &\Gamma^{\phi(x)}_{ \phi(x') g_{\mu \nu}(x'')}=\frac{1}{4}g^{\mu\nu}\delta(x,x')\delta(x,x''),\nn
&\Gamma^{g_{\lambda\tau}(x)}_{g_{\mu\nu}(x')g_{\rho\sigma}(x'')}\!\!=\delta(x,x'')\delta(x',x'')\!\left[
\frac{g^{\mu\nu}}{4}\delta^\rho_{(\lambda}\delta^{\sigma}_{\tau)}+\frac{g^{\rho\sigma}}{4}\delta^\mu_{(\lambda}\delta^{\nu}_{\tau)}\right.\nonumber\\
&\left.-\delta^{(\mu}_{(\lambda} g^{\nu)(\rho} \delta^{\sigma)}_{\tau)}
+\frac{1}{2(d-2)}\left(g_{\lambda\tau}g^{\mu(\rho}g^{\sigma)\nu}-\frac{1}{2}g_{\lambda\tau}g^{\mu\nu}g^{\rho\sigma}\right)\right]. \label{eq:Gammas}
\end{align}
The latter can be calculated in the usual way, $\Gamma^{k}_{ij} = (1/2)\cdot \mathfrak{g}^{k l} (2 \partial_{(i}\mathfrak{g}_{j)l} - \partial_{l} \mathfrak{g}_{ij})$. The presence of the connection terms will be crucial for our results - as we will show below, combined with the presence of the cubic Galileon term, they will introduce genuinely novel interactions in the gravity-scalar system leading to a new operator structure in the 1--loop effective action, unseen in the standard formalism (i.e flat field-space metric $\mathfrak{g}_{ij} = \delta_{ij}$).

The generators appearing in the second term of \eqref{eq:S_quad} are to be found by studying the gauge invariance of the classical theory. The classical action \eqref{action:0} enjoys two kind of symmetries, namely the invariance under general coordinate transformations and under Galilean transformations respectively. The latter is a global symmetry and will not contribute to the gauge fixing part of the effective action. Instead, an infinitesimal coordinate transformation, $x^\mu\rightarrow \widetilde{x}^\mu=x^\mu+\delta\epsilon^\mu(x)$, changes the two basic field variables $g_{\mu\nu}$ and $\phi$ according to 
\begin{eqnarray}
 \delta g_{\mu\nu}^{\textrm{coor}}(x)&=&\int d^n x' K^{g_{\mu\nu}(x)}{}_\lambda(x,x')\delta\epsilon^\lambda(x'), \nn
 \delta\phi^{\textrm{coor}}(x)&=&\int d^n x' K^{\phi(x)}{}_\lambda(x,x')\delta\epsilon^\lambda(x')\,,
\end{eqnarray}
where 
\begin{eqnarray}\label{generators}
 K^{g_{\mu\nu}(x)}{}_\lambda(x,x')&=&- g_{\mu\nu,\lambda}(x)\delta(x,x')-2g_{\lambda(\nu}(x)\partial_{\mu)}\delta(x,x'), \nn
 K^{\phi(x)}{}_\lambda(x,x')&=&-\partial_\lambda\phi(x)\delta(x,x'),
\end{eqnarray}
are the two generators of the transformation of the metric and scalar field respectively. 

When calculating the Gaussian piece of the action, $S_{\text{quad}}$, the background-independence of the formalism allows us to choose for simplicity a flat, Euclidean background for the metric, while for the scalar we choose a generic one as
\begin{equation}
g_{\mu \nu} = \delta_{\mu \nu} + \kappa h_{\mu \nu}, \; \; \; \phi = \bar{\phi} + \psi, \label{eq:field-expansion}
\end{equation}
with the fluctuations assumed to be small. Under (\ref{eq:field-expansion}), and using the generators \eqref{generators}, the gauge-fixing part condition $\chi_\nu$ becomes
\begin{align}
\chi_\nu=K_{i \nu} \eta^{i} = \frac{2}{\kappa}\left( \partial^\mu h_{\mu \nu} - \frac{1}{2}\partial_\nu h\right) - \omega \partial_{\nu} \phi \psi,
\end{align}
with $\omega$ a book-keeping parameter to keep track of the terms coming from the scalar-field piece in the gauge condition $\chi_\nu = 0$.

Using (\ref{eq:field-expansion}), we expand the action (\ref{action:0}) up to second order in field fluctuations and organize terms according to their order in the background scalar as\footnote{In order to ease the notation, we omit the overbar denoting the background scalar field.}
\begin{equation}
S_{\text{quad}} = S_{0}(\phi^0; \eta^i) + S_{1}(\phi^1; \eta^i) + S_{2}(\phi^2; \eta^i), \label{eq:S_quad2}
\end{equation}
whose explicit expressions are given in the Appendix. The graviton's and scalar's propagators can be read off from the $0$-th order term, and in momentum space they read as \footnote{We adopt the following momentum-space convention: $\mathcal{G}(x,y) = \int \frac{d^n p}{(2 \pi)^n} \mathcal{G}(p) \cdot e^{i p (x-y)}$.}
\begin{align}
G_{\alpha\beta\gamma\delta}(p)=&\frac{\delta_{\alpha\gamma}\delta_{\beta\delta}+\delta_{\alpha\delta}\delta_{\beta\gamma}
 -\frac{2}{d-2}\delta_{\alpha\beta}\delta_{\gamma\delta}}{2(p^2{-2\lambda})}+\nn
&\hspace{-1cm}+ (\alpha-1)\frac{\delta_{\alpha\gamma}p_{\beta}p_{\delta}
 +\delta_{\alpha\delta}p_{\beta}p_{\gamma}+p_{\alpha}p_{\gamma}\delta_{\beta\delta}+p_{\alpha}p_{\delta}\delta_{\beta\gamma}}
 {2 (p^2{-2\lambda})(p^2{-2\alpha\lambda})},  
\end{align}
\begin{align}
G(p)=\frac{1}{p^2+m_\Lambda^2}, \; \; m_\Lambda^2 = \gamma \cdot \frac{d\Lambda}{2-d}, \label{eq:Scalar-prop}
\end{align}
and $\lambda \equiv\Lambda+\gamma\Lambda\left(\frac{d-4}{4-2d}\right)$. Despite the shift-symmetry of the original action, at the perturbative level the scalar acquires an effective, mass-type interaction of the order $\sim \Lambda$, as can be also seen from (\ref{eq:S_0}). Its origin is identified in the term $\Gamma^{g}_{\phi \phi} \partial_{g} S$ of the connection--dependent piece of (\ref{eq:cov-action}) (see also (\ref{eq:Gammas})), and is absent for a flat field-space metric ($\gamma = 0$). As we will see shortly, it will have an important impact on the 1--loop structure of the effective action.

In a perturbative approach, the terms $S_1$ and $S_2$ in (\ref{eq:S_quad2}) are treated as small perturbations. Expanding the exponential in (\ref{action:effective0}), the effective action at second order reads 
\begin{align}
\Gamma^{\text{1--loop}} \simeq \Braket{S_2(x,x)} - \frac{1}{2}\Braket{S_{1}(x) \cdot S_{1}(x')} + \mathcal{O}(\bar{\phi}^3)\,, \label{eq:Eff-action2}
\end{align}
where the brackets, by means of Wick's theorem, lead to UV-divergent loop-integrals of  propagator products, and $\Braket{h_{\mu \nu}(x) \psi(x')} = 0$. Truncating (\ref{eq:Eff-action2}) up to second-order in the background scalar implies that the cubic term in (\ref{action:0}) will not get itself renormalized, though it will have a non--trivial effect on lower-order operators. At this stage we should notice that, the gauge fixing sector of the theory introduces the associated ghost fields through the Faddeev--Popov procedure. It can be shown that they contribute to quartic order in the background field \cite{Mackay:2009cf}, and hence we omit them. 

The second term on the r.h.s of (\ref{eq:Eff-action2}) introduces graviton--scalar and scalar-scalar derivative interactions of the form $\partial^n \Braket{  h_{\mu \nu} h_{\kappa \lambda}} \cdot \partial^m\Braket{\psi \psi}$, $\partial^n \Braket{\psi \psi}\cdot \partial^m \Braket{\psi \psi}$ with $n+m = 0, \cdots 6$. \footnote{Their precise expressions can be straightforwardly derived using the explicit relations (\ref{S1:explicit})--(\ref{S2:explicit}).}, while the first one is of the tadpole-form. It is easy to see from (\ref{eq:Eff-action2}) in combination with (\ref{S1:explicit})--(\ref{S2:explicit}) that the graviton-graviton interactions will appear only in the tadpole term, and as it turns out after evaluating the trace integrals, they do not contribute to the pole-part of the effective action.

Working in momentum space, we evaluate the momentum traces in (\ref{eq:Eff-action2}) using the gauge-invariant scheme of dimensional regularization, which is sensitive to the logarithmically-divergent part of the integral. Since we are interested in the pole part of the effective action, when working in momentum space, we expand the integrand in powers of $q^{-1}$ for large momenta $q$, symmetrizing the numerator according to
\begin{align}
q_\mu q_\nu=q^2 \frac{g_{\mu\nu}}{d}, \; q_\mu q_\nu q_\rho q_\sigma=\frac{q^4}{d(d+2)}\cdot 3\cdot g_{(\mu\nu}g_{\rho\sigma)},
\end{align}
and similar completely symmetrized expressions for higher orders. The expansion truncates at $q^{-d}$, which' integration leads to the diverging factor 
$a_L\stackrel{\tiny{d\rightarrow4}}{=}-\frac{1}{8\pi^2 (d-4)}$ in four dimensions. After a tedious calculation, we
obtain the divergent part of the effective action in $d=4$ as
\begin{widetext}
\begin{align}
\Gamma^{\text{1--loop}} \stackrel{\tiny{\alpha\rightarrow0}}{=} a_L \cdot \int d^4 x \Bigg\{ & -\frac{1}{16M^6}\phi\Box^{(4)}\phi    \; +\frac{5m_\Lambda^2}{8M^6}\phi\Box^{(3)}\phi  \; 
+ \phi\Box^{(2)}\phi \left[ \frac{\kappa^2}{4} \cdot \left( \alpha\gamma+\frac{3\gamma^2}{4}-\frac{3\gamma}{2}-\frac{\alpha\gamma^2}{4}-\omega-\frac{\gamma\omega}{2} \right) -\frac{15m_\Lambda^4}{8M^6}\right]\nn
& + \partial_\mu\phi\partial^\mu\phi \cdot \frac{\kappa^2}{2} \cdot \left[\frac{\gamma m_\Lambda^2}{8} -  \lambda\omega^2  -  \omega m_\Lambda^2 +2\alpha\lambda \omega+\frac{\alpha m_\Lambda^2}{2}- \alpha^2\lambda\right]\Bigg\}. \label{Eff-action}
\end{align}
\end{widetext}
{\it The effective action (\ref{Eff-action}) constitutes the main result of this paper.} The limit $\alpha \to 0$ is understood here, corresponding to the gauge--independent result, but we kept $\alpha \neq 0$ to highlight the gauge--dependence of the various terms. Most importantly, the parameter $\gamma$ tracks all terms stemming from the field-space connection and should be set to $\gamma=1$. Note finally that in $d=4$ dimensions, the parameter $\lambda$ reduces to the cosmological constant $\Lambda$.
We also stress the remarkably interesting fact associated with the lack of terms $\sim \alpha^{-n}$; such terms, appear in the different stages of the calculation, however they cancel each other when all the contributions to the effective action are summed, ensuring the finiteness of the action as $\alpha\rightarrow0$. 

The diverging part of the effective action (defined through (\ref{Eff-action}) as $d \to 4$) is unphysical and should be eliminated. The renormalization procedure requires that the diverging piece is cancelled through the introduction of appropriate counterterms in the original, bare action. The latter can be trivially read-off from (\ref{Eff-action}).

We now discuss the particular form of the 1--loop result (\ref{Eff-action}). It exhibits novel features as a by-product of the quantum--gravitational fluctuations and the requirement of background--independence of the calculation respectively. Let us first look at the renormalization of the kinetic term, where it is well--known that at 1--loop order scalar loops do not account for a correction to the scalar's kinetic term (i.e wave-function renormalization is zero). This is exactly the case of the kinetic term in (\ref{Eff-action}), where all terms result from the graviton-scalar interactions in the second term of the r.h.s of (\ref{eq:Eff-action2}). Notice that all terms are Planck--suppressed through the coefficient $\kappa^2 \sim 1/M_{p}^2$, as expected. 

On the other hand, the higher--derivative operators $\sim m^{2}_{\Lambda}$, stem from pure scalar derivative interactions in (\ref{eq:Eff-action2}), and they are intimately related to the coupling to gravity, through the cosmological constant $\Lambda$. Most importantly, they vanish for a flat field-space metric ($\gamma= 0$ and $m^{2}_{\Lambda} = 0$), and in this sense, they are invisible in the standard background-field formalism. What is more, they also vanish as $M \to \infty$, corresponding to the limit where the cubic derivative interaction in the effective action decouples. In this sense, their presence is also closely related to the cubic Galileon interaction in the bare action.

It is interesting to notice that the operators $ \phi\Box^{(3)}\phi, \phi\Box^{(2)}\phi$ would also appear within a dimensionful-cut-off regularization scheme, representing quartic and quadratic divergences \cite{Brouzakis:2013lla} ($\sim \Lambda_{\text{cut--off}}^4$ and $\Lambda_{\text{cut--off}}^2$ respectively \footnote{Notice that, cut--off regularization does not in principle respect the gauge symmetry of the theory.}), also in the absence of gravitational interactions, while in the usual background-dependent calculation within dimensional regularization only the first term in (\ref{Eff-action}) would appear, representing the standard logarithmic divergence. The crucial difference in (\ref{eq:Eff-action2}) generating the novel operator structure is indeed the mass scale introduced through $m^{2}_{\Lambda}$ in the Galileon's propagator (see (\ref{eq:Scalar-prop})), which is essentially the scale of the  vacuum energy $\Lambda$. In a pure scalar-field theory, the zero-point fluctuations, represented by $\Lambda$, correspond to a physically-irrelevant infinite term, which can be always shifted away, however, this is no longer true in the presence of gravity. One expects therefore, that in the absence of gravity, the scalar propagator would indeed remain massless. 
Notice also that, in the limit of a canonical scalar-field theory, $M \to \infty$, (\ref{eq:Eff-action2}) is consistent with the results quoted in ref. \cite{Mackay:2009cf}.

The emergence of the higher--order operators in the scalar sector suggests that they need to be included in the original bare action for consistency of the renormalization procedure \cite{Collins:105730}. Obviously, in that case, the second-order character of the equations will be lost and the stability of the theory's dynamics endangered through the well--known Ostrogradsky instability associated with theories of order higher than two \cite{wood}. What is more, these (gravitationally-induced) operators, resulting from the Galileon running in loops, are not Planck-mass suppressed, in contrast to the corrections originating from graviton loops. This means that, if these operators are indeed included in the original (bare) action, care has to be taken for understanding how they affect the theory's classical dynamics, particularly in scenarios where the Galileon plays the role of the dark energy or the inflaton field \cite{Chow:2009fm,Silva:2009km,Kobayashi:2010cm,DeFelice:2010nf}. It is further important to understand how they might possibly affect the effective operation of the Vainshtein mechanism and the radiative stability of the theory along the lines of refs \cite{Kaloper:2014vqa,Pirtskhalava:2015nla}. We believe that these issues deserve their own study and that our results will provide the motivation in this direction.

\section{Summary} \label{sec:Discussion}
Galileon scalar--field theories appear as an effective field--theory limit of different, distinct extensions of the standard paradigm of gravity. Their uniqueness lies in their higher--order derivative interactions, while retaining the second--order character of the classical dynamics, their remarkable renormalization properties, as well as their ability to account for the early-- or late--time acceleration of the Universe. The non--linear derivative configurations of the scalar underline the Vainshtein screening mechanism which acts to suppress the coupling to matter sufficiently close to the source, making in this context any calculation of perturbative corrections sensitive to the background--choice.

In this work, we derived for the first time the 1--loop effective action for the cubic Galileon theory coupled to quantum--gravitational fluctuations, within the field-reparametrization-invariant framework of DeWitt and Vilkovisky. Our analysis revealed novel interactions, induced through the interplay of the non--trivial derivative structure of the original theory and the coupling of the Galileon field to gravity. Field--reparametrization invariance required that the Galileon field acquired an effective mass at the perturbative level, although the original theory was shift symmetric. The latter fact was crucial for the new operator structure revealed in the effective action, unnoticed in previous calculations. We further showed that quantum--gravitational effects provide the sole contribution to the wave--function renormalisation of the scalar field as can be seen from the effective action (\ref{Eff-action}), and as expected from well--known results in simpler setups. The effective action (\ref{Eff-action}) defines the counterterms to renormalize the bare action at 1--loop, from which one can formally extract the beta functions of the theory. It further suggests the new (derivative) interactions which need to be added in the original action for consistency of the renormalization program. 

In this relatively simple, yet non--trivial theory setup we considered, our results highlighted the significance of the field--reparametrization character of the approach, important for theories sensitive to the choice of background or gauge. 

We hope that this work will provide the motivation to further explore the coupling to gravity of more general theories within the Galileon family in this context, as well as the effect of the associated higher--order corrections for the classical and quantum phenomenology of the theory.

\section{acknowledgements}
VV is supported by the FCT-Portugal grant SFRH/BPD/77678/2011. IDS is supported by FCT under the grant SFRH/BPD/95204/2013, and further acknowledges UID/FIS/04434/2013 and the project FCT-DAAD 6818/2016-17. IDS is thankful to Tony Padilla, Iggy Sawicki and Dimitri Skliros for enlightening discussions. VV would like to thank Nino Flachi for useful discussions.

\begin{widetext}
\vspace{-0.3cm}
\section*{Appendix} \label{sec:Appendix}
\vspace{-0.2cm}
Here we present the explicit form of the terms appearing in (\ref{eq:S_quad2}), corresponding to the quadratic action in field fluctuations. We remind that the book-keeping parameter $\gamma$  tracks the effect of the field-space connection, and $\phi_{\mu}, \psi_{\mu} \equiv \partial_{\mu} \phi, \partial_{\mu} \psi$. 
\begin{align}
S_0&=\int d^d x \cdot \left\{\frac{1}{2}\delta^{\mu\nu}\psi_\mu\psi_\nu-\frac{1}{2}h^{\mu\nu}\Box h_{\mu\nu}+\frac{1}{4}h \Box h+\left(\frac{1}{\alpha}-1\right)
 \left(\partial^\mu h_{\mu\nu}-\frac{1}{2}\partial_\nu h\right)^2+\right.\nn
 &\left.\hspace{1cm}{+\Lambda\left(\frac{h^2}{2}
 -h^{\mu\nu}h_{\mu\nu}\right)\left[1-\frac{\gamma}{2}\left(\frac{d-4}{d-2}\right)\right]}{
 -\gamma\Lambda\frac{d}{2d-4}\psi^2}\right\} \label{eq:S_0}\\ 
S_1&=\!\int  d^d x \cdot \left\{\frac{1}{M^3}\left(\frac{1}{2}\Box\phi\psi_\mu\psi^\mu+\phi_\mu\psi^\mu\Box\psi\right)
+\frac{\kappa}{2} \phi_\mu h\psi^\mu 
-\kappa h^{\mu\nu}\psi_\mu\phi_\nu
 +\frac{\kappa\gamma}{4}h\psi\Box\phi-\frac{\kappa\omega}{\alpha}\left(\partial^\lambda h_{\lambda\nu}-\frac{1}{2}\partial_\nu h\right)\partial^\nu\phi~\psi\right\} \label{S1:explicit}\\
 S_2&= \kappa^2\int  d^d x \cdot \left\{\frac{1}{2}X\left(\frac{h^2}{4}-\frac{1}{2}h_{\mu\nu}h^{\mu\nu}\right)-\frac{1}{4}hh^{\mu\nu}\phi_\mu\phi_\nu+\frac{1}{2}h^\mu{}_\alpha
 h^{\alpha\nu}\phi_\mu\phi_\nu+\right.\nonumber\\
 &+\frac{1}{M^3}\left[\frac{h}{2}\Big(\Box\phi \phi_{\mu}\psi^\mu+X\Box\psi\Big)-h^{\mu\nu}\psi_\mu\phi_\nu\Box\phi +\phi_\beta~\psi^\beta\left(-h^{\mu\nu}\phi_{\mu\nu}-\frac{1}{2}\phi^{\rho}
\left(2\partial_\mu h^{\mu}{}_{\rho}-\partial_\rho h\right)\right)\right.\nn
&\left.-\frac{1}{2}h^{\mu\nu}\phi_\mu\phi_\nu\Box\psi+X\left(-h^{\mu\nu}\psi_{\mu\nu}-\frac{1}{2}\psi^{\rho}
\left(2\partial_\mu h^{\mu}{}_{\rho}-\partial_\rho h\right)\right)\right]+\frac{\gamma}{M^3}\frac{h}{4}\psi(\Box\phi\Box\phi-\partial^\alpha\partial^\mu\phi\partial_\alpha\partial_\mu\phi)+\nn
&\left.+\gamma\left[-\frac{1}{16}\psi^2\phi_\mu\phi^\mu
+\frac{1}{16} h_{\mu\nu} h^{\mu\nu} \phi_{\lambda} \phi^{\lambda} - \frac{1}{32} h^{\mu}{}_{\mu} h^{\nu}{}_{\nu} \phi_{\lambda} \phi^{\lambda}-\frac{1}{4}h_{\lambda}{}^{\nu} h_{\mu\nu} \phi^{\lambda} \phi^{\mu} +\frac{1}{8} h_{\lambda\mu} h^{\nu}{}_{\nu} \phi^{\lambda} 
\phi^{\mu}\right]+\frac{\omega^2}{4\alpha}\partial^\mu\phi\partial_\mu\phi~\psi^2\right\} \label{S2:explicit}
\end{align}
\end{widetext}

\bibliographystyle{utcaps}
\bibliography{VG.bib}

\end{document}